\documentclass[a4paper,11pt]{article}
\usepackage{pos}

\usepackage[english]{babel}  
\usepackage[utf8]{inputenc} 
\usepackage{hyperref}    
\usepackage[separate-uncertainty=true,multi-part-units=single]{siunitx}
\usepackage{subcaption}
\usepackage{diagbox}
\usepackage{makecell}
\usepackage{verbatim} 
\usepackage{natbib}
\usepackage{xcolor}

\title{Towards a Cosmic-Ray Energy Scale with the \\Auger Engineering Radio Array}
\ShortTitle{Towards a Cosmic-Ray Energy Scale with AERA}

\manuallySeparateAuthors 
\author*[a,b]{Max Büsken}
\author[c, \dag]{ for the Pierre Auger Collaboration\notes{\note{Full author list at \url{http://www.auger.org/archive/authors_2024_06.html}.}}} 
\affiliation[a]{Institute for Experimental Particle Physics (ETP), Karlsruhe Institute of Technology (KIT),\\ Hermann-von-Helmholtz-Platz 1, 76344 Eggenstein-Leopoldshafen, Germany}
\affiliation[b]{Instituto de Tecnologías en Detección y Astropartículas (CNEA, CONICET, UNSAM),\\ Av. General Paz 1555 (B1630KNA), San Martín, Buenos Aires, Argentina}
\affiliation[c]{Observatorio Pierre Auger, Av. San Martín Norte 304, 5613 Malargüe, Argentina}

\emailAdd{spokespersons@auger.org}

\abstract{Radio detection of cosmic-ray (CR) induced extensive air showers with digital antenna arrays is a matured technique by now. At the Pierre Auger Observatory, the Auger Engineering Radio Array (AERA) has been measuring air-shower signals in conjunction with the particle detectors of the surface detector (SD) for over ten years. For an absolute determination of the CR energy with the Auger baseline detectors, the shower size estimator from the SD is calibrated with the energy scale of the fluorescence detector (FD). However, AERA has an independent access to the energy scale through the reconstructed radio signals. The hybrid detectors at the Pierre Auger Observatory offer the unique opportunity to compare the two independent energy scales. In this contribution, we present our envisaged methodology for cross-checking the agreement between the energy scales of the FD and AERA using hybrid SD-AERA shower data and simulations. We show individual steps of our radio signal reconstruction and highlight the key ingredients for calibrated energy measurements.}

\FullConference{10th International Workshop on Acoustic and Radio EeV Neutrino Detection Activities (ARENA2024)\\
11-14 June 2024\\
The Kavli Institute for Cosmological Physics, Chicago, IL, USA\\}

\begin{document}
\maketitle

\section{Introduction}
\label{sec:1_Introduction}

In the field of CR research, many features have been studied to great precision but some open questions about the results found remain. At the highest energies, detecting CRs is only feasible in an indirect way by observing the signals from the collision of those particles with a medium, for example, creating extensive air showers from the interaction with the atmosphere. Radio signals from such air showers have been studied in various aspects already and their detection and the reconstruction of CR properties have come a long way. The maturing radio technique is contributing to CR research with growing importance \cite{Huege_2016, Schroeder_2017}.

A key observable is the energy of the primary CR particle whose accurate determination is crucial to interpret the results. In this regard, the fluorescence technique is currently used to determine the CR energy scale. However, some of the results from the two CR observatories measuring at ultra-high energies, the Telescope Array and the Pierre Auger Observatory, show differences that are yet to be understood \cite{Tsunesada_2023}. Part of an explanation could be a possible difference between their CR energy scales.

With the well-calibrated AERA detector, a thoroughly developed event reconstruction, a high-quality dataset and precise simulations, we aim to cross-check the CR energy scale of the Pierre Auger Observatory. The radio detection technique brings a few intrinsic advantages to this task that set it apart from the fluorescence technique: there is nearly no absorption or scattering of the radio signal in the atmosphere and there is no sign of significant or relevant ageing of radio detectors \cite{deAlmeida_2023}. Also, radio detectors can be operated around the clock.

Through the concept of the presented analysis, the radio energy scale is set by Monte Carlo simulations of the electromagnetic component of extensive air showers in combination with a microscopic, parameter-free description of the air shower's radio emission that is derived from first-principles classical electrodynamics.


\section{The Pierre Auger Observatory}
\label{sec:2_Auger}

Located in the Pampa Amarilla in the west of Argentina, close to the city of Malargüe, the Pierre Auger Observatory is the largest facility for the detection of ultra-high energetic cosmic rays world-wide. The total area covered by detectors is over \SI{3000}{km^2}. A special feature of the Pierre Auger Observatory is its hybrid structure with multiple different detection techniques employed at the same site, allowing for the simultaneous observation of the same showers and cross-detector analyses. The ongoing AugerPrime detector upgrade will allow to deepen and extend the research done at the observatory \cite{Auger_Status_Upgrade_2020}.

Two of the main components of the Pierre Auger Observatory are the FD and the water-Cherenkov detector of the SD. While the former provides a near-calorimetric measurement of the air-shower energy, which can be converted to the primary CR's energy through correction for the missing so-called invisible energy, the SD measurement cannot directly access the energy of the primary CR. However, hybrid detection of showers with both the FD and the SD allow to cross-calibrate the SD energy estimator with the CR energy from the FD. This way, the SD inherits the CR energy scale from the FD. \cite{Dawson_2019}

The SD has a standard detector spacing of \SI{1500}{m} but in a sub-array, a denser spacing of \SI{750}{m} is employed, allowing to measure showers at lower energies. In this work, we use the \SI{750}{m} sub-array of the SD.

\begin{figure}
  \begin{subfigure}[c]{0.49\textwidth}
    \center
    \includegraphics[width=0.72\textwidth]{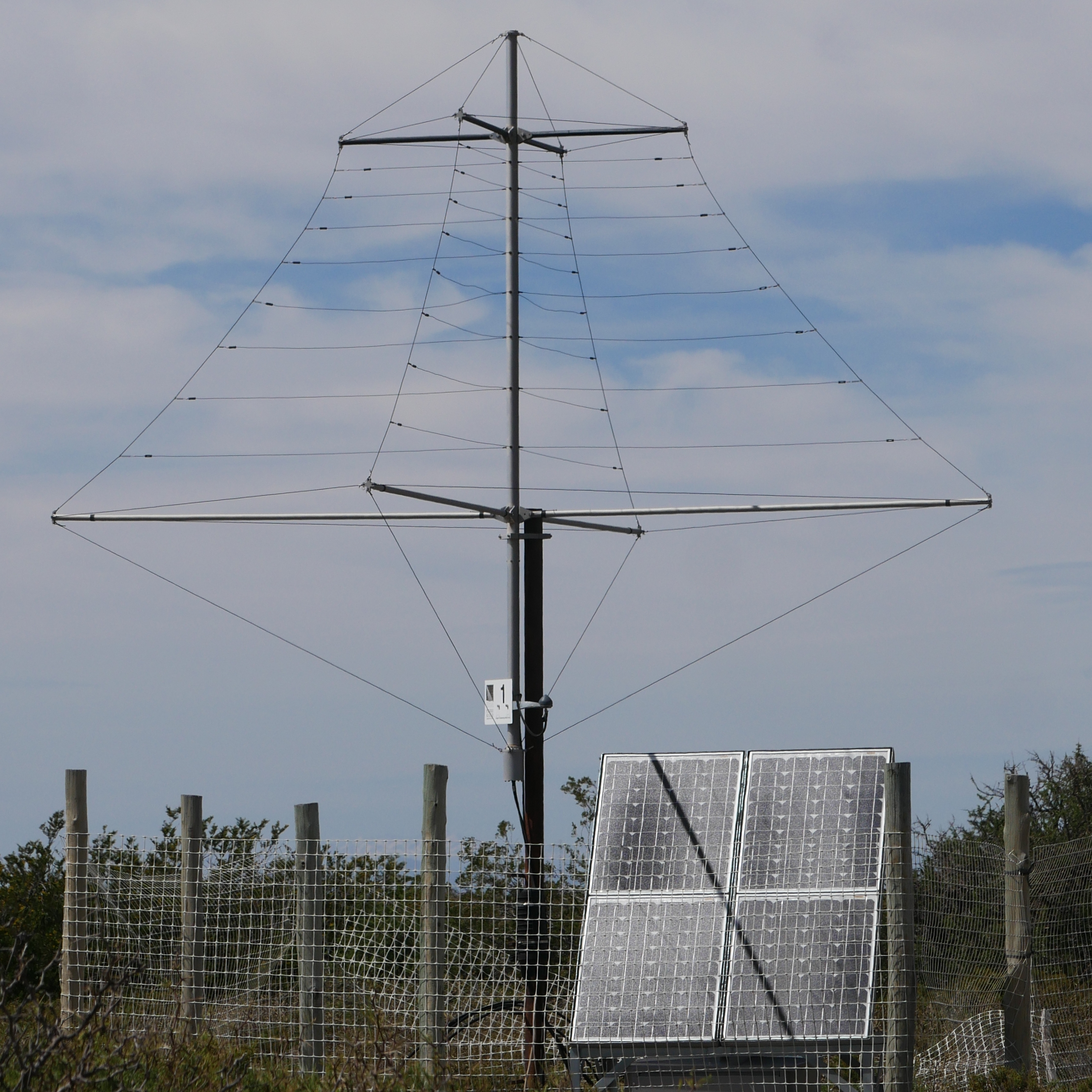}
    \subcaption{LPDA}
  \end{subfigure}\hfill
  \begin{subfigure}[c]{0.49\textwidth}
    \center
    \includegraphics[width=0.72\textwidth]{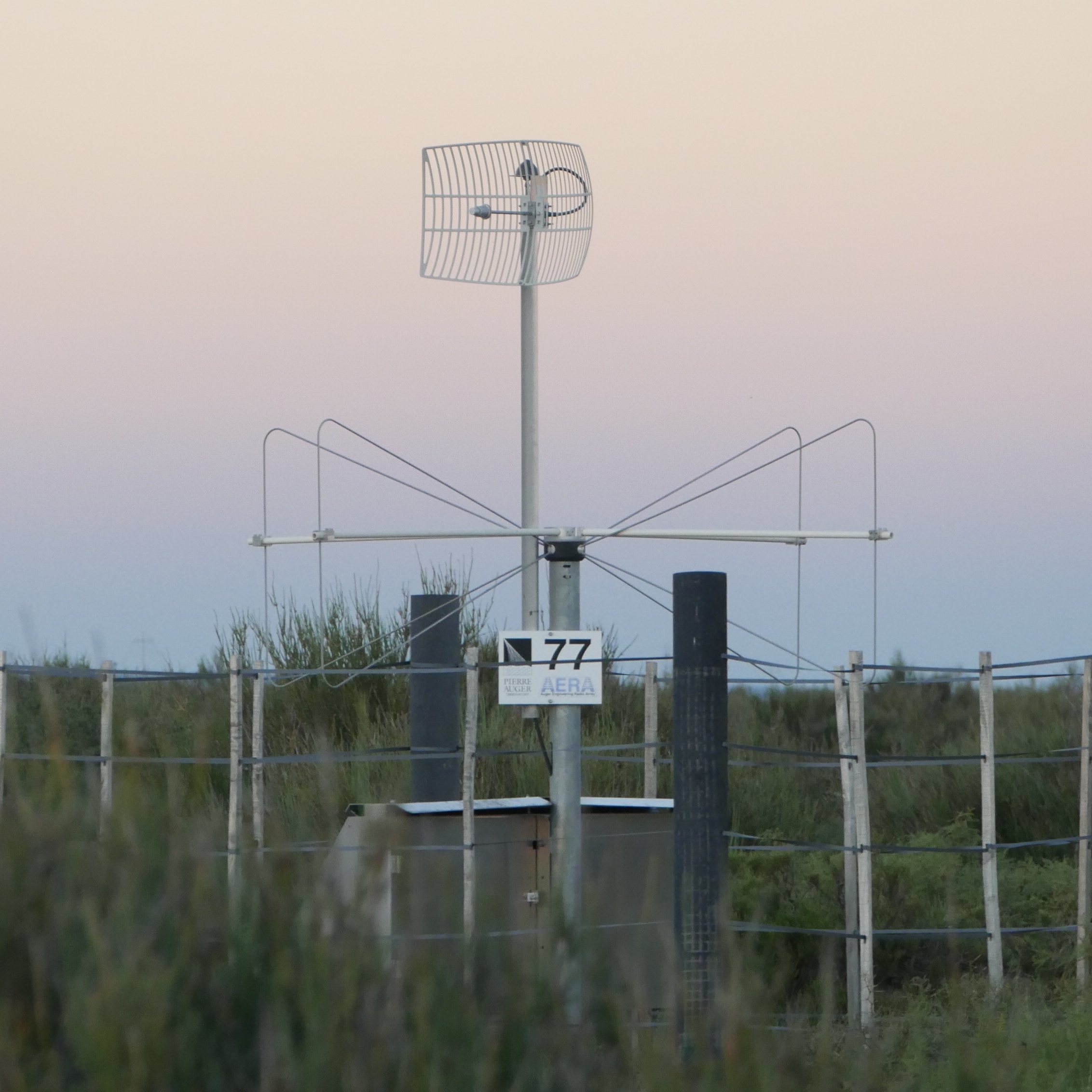}
    \subcaption{Butterfly}
  \end{subfigure}
  \caption{Two AERA stations in the field.}
  \label{fig:stations}
\end{figure}

Within the SD-750 array, 153 antenna stations of the Auger Engineering Radio Array are located that measure air-shower radio pulses in the frequency band from 30 to \SI{80}{MHz} \cite{Pont_2023}. Out of the 153 stations, 24 are equipped with log-periodic dipole antennas (LPDA) while at the other stations, Butterfly antennas are installed. Pictures of an LPDA station and a Butterfly station are shown in Fig.\ \ref{fig:stations}. The stations are further separated between two different triggering systems, self-triggered and externally triggered. We only use the subset of 113 stations that are able to receive an external trigger from the SD.

A major component of the AugerPrime upgrade will be the new Radio Detector (RD) comprised of short aperiodic loaded loop antennas (SALLA) installed on each of the SD stations. The RD will extend the air-shower detection of the Pierre Auger Observatory to very inclined showers and allow new and ongoing CR studies to be done with unprecedented precision \cite{Schlüter:2021rh}.

\section{Ingredients of the radio event reconstruction}
\label{sec:3_radio_reconstruction}

The reconstruction of air-shower properties from radio signals measured by AERA has been developed over a long period within the \emph{Offline} software framework of the Pierre Auger Collaboration \cite{Radio_Offline}. Many ingredients are combined to ensure an accurate reconstruction with a good absolute calibration, some of which are highlighted in the following.

The signal responses of all components of the AERA readout electronics were measured in the lab. The directional response pattern of the LPDA antenna was simulated with the NEC-2 software and rescaled with drone calibration measurements \cite{LPDA_calibration_2017}. For the Butterfly stations, a simulated response pattern is used that will be validated with measurements from a recent drone calibration campaign \cite{Reuzki_2024_ARENA}. All responses are incorporated in the signal reconstruction.

Amplifier gains of the readout electronics are temperature dependent causing seasonal variations in the measurements. A clear correlation was found and is corrected for in the event reconstruction using temperature information from a nearby weather station. After temperature correction, a Galactic calibration is applied to the signal spectra. The calibration is done per station, per month, per antenna channel and per \SI{1}{MHz} frequency bin. Its systematic uncertainty is estimated to be around 6\% \cite{Correia_2024_ARENA}.

For each station, the local radio energy fluence is currently estimated using a simple noise subtraction approach. However, we plan to improve this step by employing a new signal estimation method based on Rice distributions \cite{Martinelli_2024}.

Lastly, we fit a lateral distribution function (LDF) based on the two main emission mechanisms, called \texttt{GeoCeLDF}, to the energy fluence estimates of all stations with a signal \cite{GeoCeLDF_2019}. The integral over the fitted LDF returns the radiation energy, $E_\text{rad}$, which denotes the energy deposited on the ground in the form of radio waves in the frequency band of AERA and which is directly related to the electromagnetic energy of the air shower \cite{PhysRevLett.116.241101, PhysRevD.93.122005, Glaser_2016}. The LDF has been validated successfully with a resolution of the radiation energy of 4\%.

There are several additional intermediate steps involved in cleaning and preparing the radio signals at different levels of the reconstruction. For example, each AERA station is continuously being monitored and, in case of malfunction, an entry is made into a \emph{bad period} database \cite{Pont_2021}. This database is queried during event reconstruction to reject malfunctioning stations. In a similar way, a different database is filled with permanent measurements of the atmospheric electric field at AERA. Since large electric fields in the presence of thunderstorms significantly alter the radio signal at ground, events recorded during thunderstorm periods are flagged and discarded from the analysis \cite{Gottowik_2021}.


\section{Comparing the energy scales from FD and AERA}
\label{sec:5_Analysis_logic}

Most of the scientific results yielded by the Pierre Auger Collaboration involve CR energy measurements and the interpretability of the results therefore depends on an accurate determination of the FD energy scale. Currently, the systematic uncertainty of this energy scale is estimated to be 14\% with the largest contribution coming from the photometric calibration of the telescopes \cite{Dawson_2019}. At the highest energies, the only other CR observatory is Telescope Array (TA), whose fluorescence energy scale has a systematic uncertainty of 21\% \cite{AbuZayyad_2015}. 

There are differences in individual results from TA and the Pierre Auger Collaboration, e.g.\ regarding the CR energy spectrum. While some of them can be explained by differences in the FD energy reconstruction, others cannot \cite{Tsunesada_2023}. Therefore, it is essential to reduce the systematic uncertainties of the respective energy scales. Our goal is to cross-check the Auger energy scale with the independent radio energy scale, which could help to reduce the systematic uncertainty of the former.



\begin{figure}
	\centering
	\includegraphics[width=0.92\textwidth]{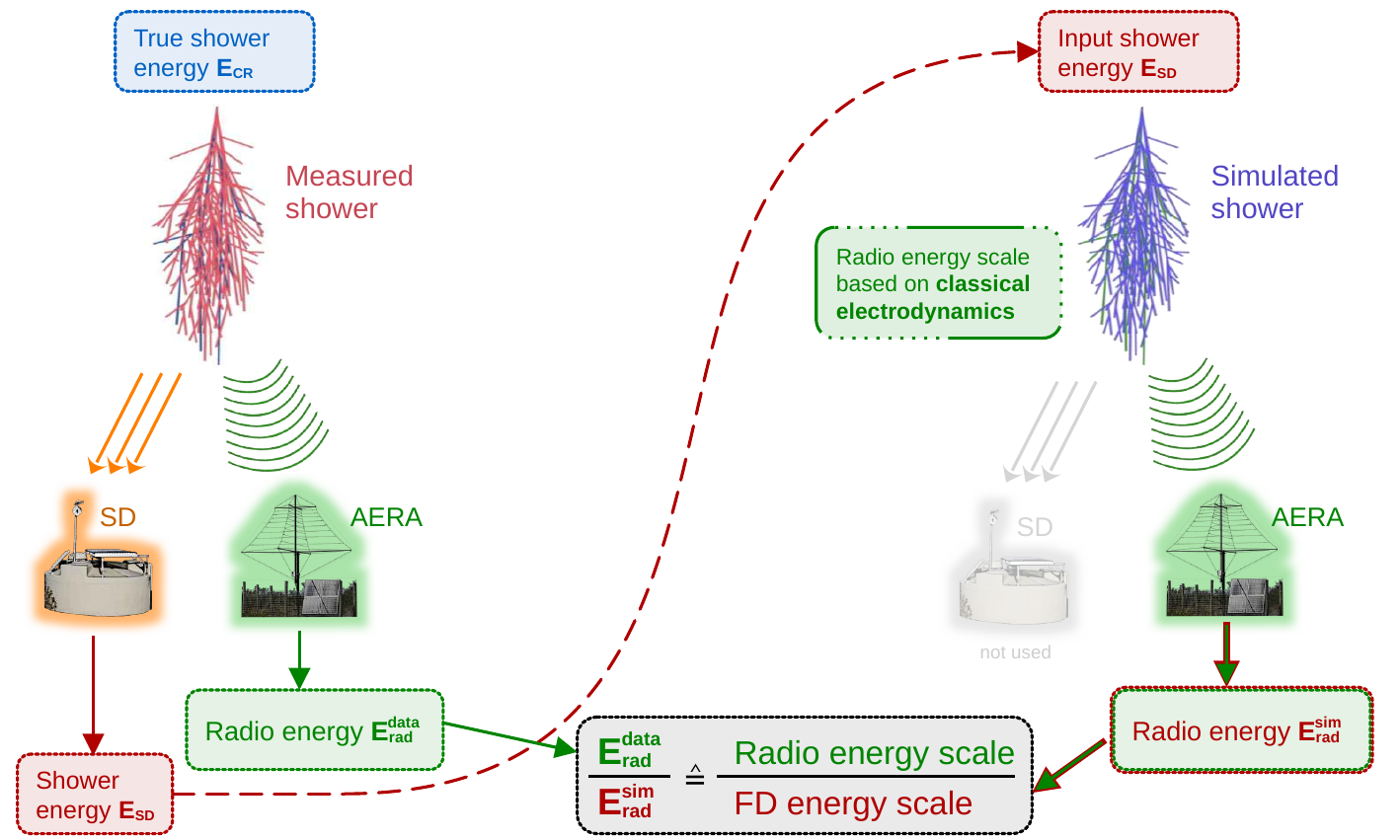}
	\caption{Proposed logic for event-by-event comparison of reconstructed measured showers with matched simulated showers. The green color indicates radio measurements and the red color denotes observables that are on the FD energy scale.}
	\label{fig:Analysis_logic}
\end{figure}

We aim to cross-check the FD energy scale by quantifying its agreement with the radio energy scale provided through AERA. Since the statistics for hybrid FD-AERA events are too small, we use hybrid SD-AERA events as the SD is calibrated to the FD energy scale. Our approach for the analysis is an event-by-event comparison of measured showers and matched simulations. The logic is illustrated in Fig.\ \ref{fig:Analysis_logic} which shows a measured hybrid event to the left and the corresponding simulation to the right. 

A measured shower is reconstructed individually from the SD and AERA signals. The reconstruction of the radio signals yields an estimate for the radiation energy, $E^\text{data}_\text{rad}$, that is on the radio energy scale. In addition, the SD reconstruction provides an estimate for the total energy of the primary cosmic ray, $E_\text{SD}$, that is on the FD energy scale. We use $E_\text{SD}$ to calculate the energy input of a shower simulation with CORSIKA 7.7550 \cite{CORSIKA} and CoREAS 1.4 \cite{CoREAS}. We subtract the invisible energy fraction described in \cite{Auger_2019_invisbleEnergy}, which was added in the reconstruction, and obtain the calorimetric energy of the shower. Then, we take a parameterization of the invisible energy fraction in the hadronic interaction model that we use in CORSIKA, Sibyll 2.3d \cite{Sibyll2.3d}, and calculate the required simulation input energy to be able to simulate a shower with that same calorimetric energy. This way, the radio signals of the measured and simulated shower are matched as close as possible and they can be compared event-by-event. 

The simulation is run with $10^{-6}$ thinning and optimized weights \cite{Corsika_Thinning} and the \texttt{STEPFC} parameter set to 0.05 to ensure a high accuracy of the radio-signal prediction \cite{Gottowik_2018}. We create realistic simulation conditions by using a grid with the actual AERA station layout and by passing an atmosphere provided by the Global Data Assimilation System (GDAS) that is taken from the location of AERA and from the time of the measured event within a resolution of \SI{3}{h} \cite{GDAS_LOFAR}. Each simulation is run twice, once with proton and once with iron as primaries.

All simulations are reconstructed with the Offline analysis framework in the same way as the measured events so that potential biases introduced during signal cleaning and processing affect data and simulation reconstructions in the same way. At the beginning of the simulation reconstruction, measured noise from the time of the event is added per station to the simulated signals. This reconstruction yields an estimate $E^\text{sim}_\text{rad}$ which is on the FD energy scale. By comparing this estimate to $E^\text{data}_\text{rad}$ of the same event, we will be able to directly probe the agreement of the FD and radio energy scales.

\section{Hybrid SD-AERA dataset}
\label{sec:6_Dataset}

The hybrid SD-AERA dataset used in this analysis contains 912 events from May 2013 until April of 2019 with zenith angles up to \SI{55}{\degree}. Those events pass standard SD quality cuts including a lower cosmic-ray energy cut at \SI{3e17}{eV} that guarantees full detection efficiency. We require at least five stations with a signal\footnote{A station is considered a signal station if its signal-to-noise ratio (SNR) fulfills $\text{SNR} = f_\text{signal}^2/f_\text{noise}^2 > 10$, where $f_\text{signal}$ and $f_\text{noise}$ are calculated as the Hilbert envelope amplitudes from signal and noise, respectively.} as well as a successful \texttt{GeoCeLDF} fit.

\begin{figure}
  \begin{subfigure}[c]{0.4\textwidth}
    \center
    \includegraphics[width=0.95\textwidth]{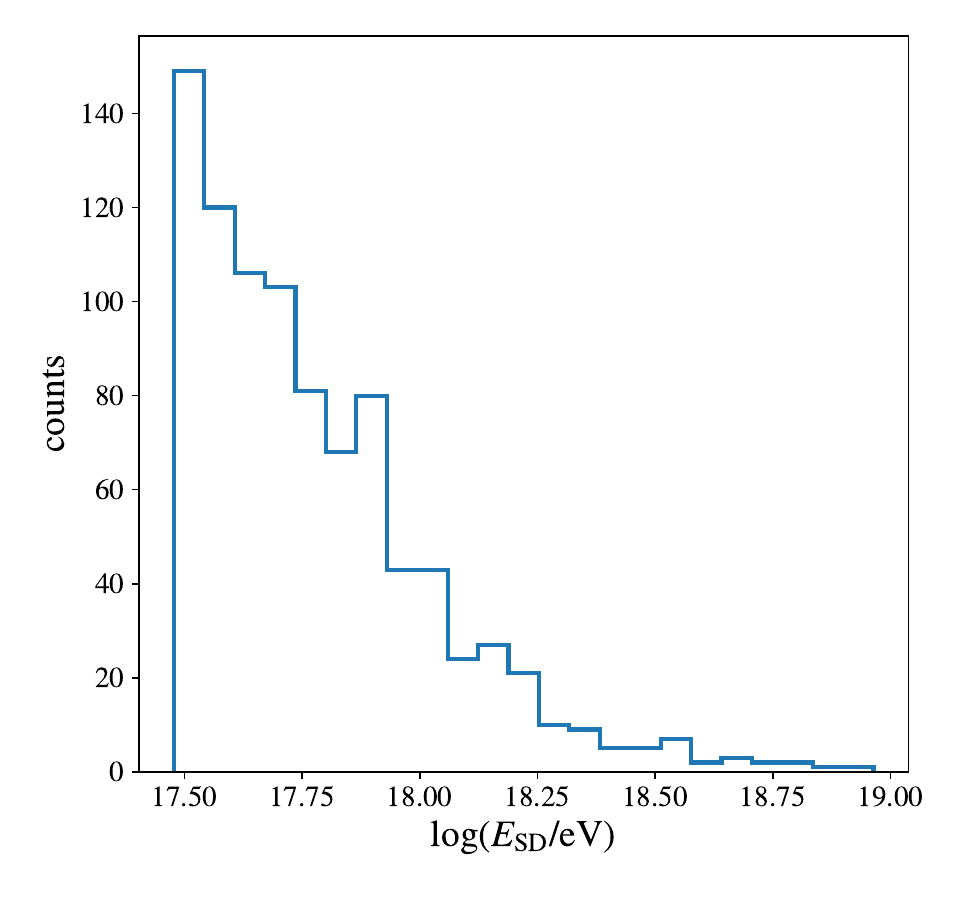}
    \subcaption{Energies}
  \end{subfigure}\hfill
  \begin{subfigure}[c]{0.6\textwidth}
    \center
    \includegraphics[width=0.95\textwidth]{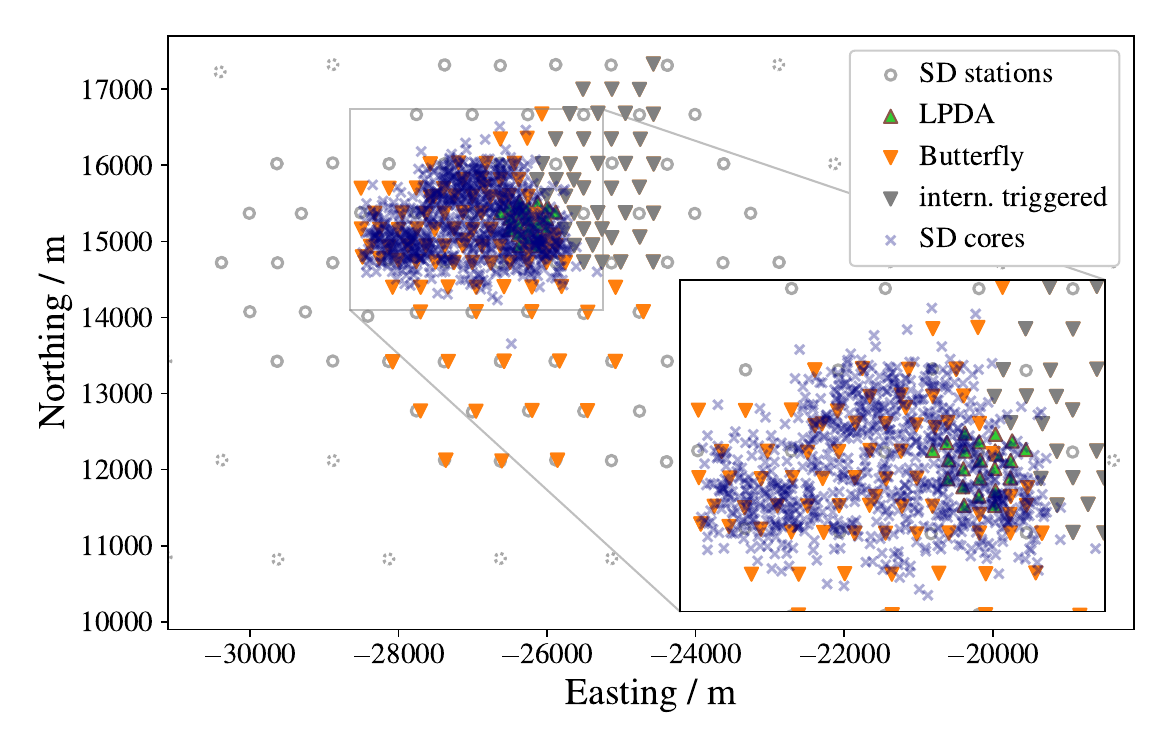}
    \subcaption{Shower cores}
  \end{subfigure}
  \vspace{5pt}
  \begin{subfigure}[c]{0.5\textwidth}
    \center
    \includegraphics[width=0.95\textwidth]{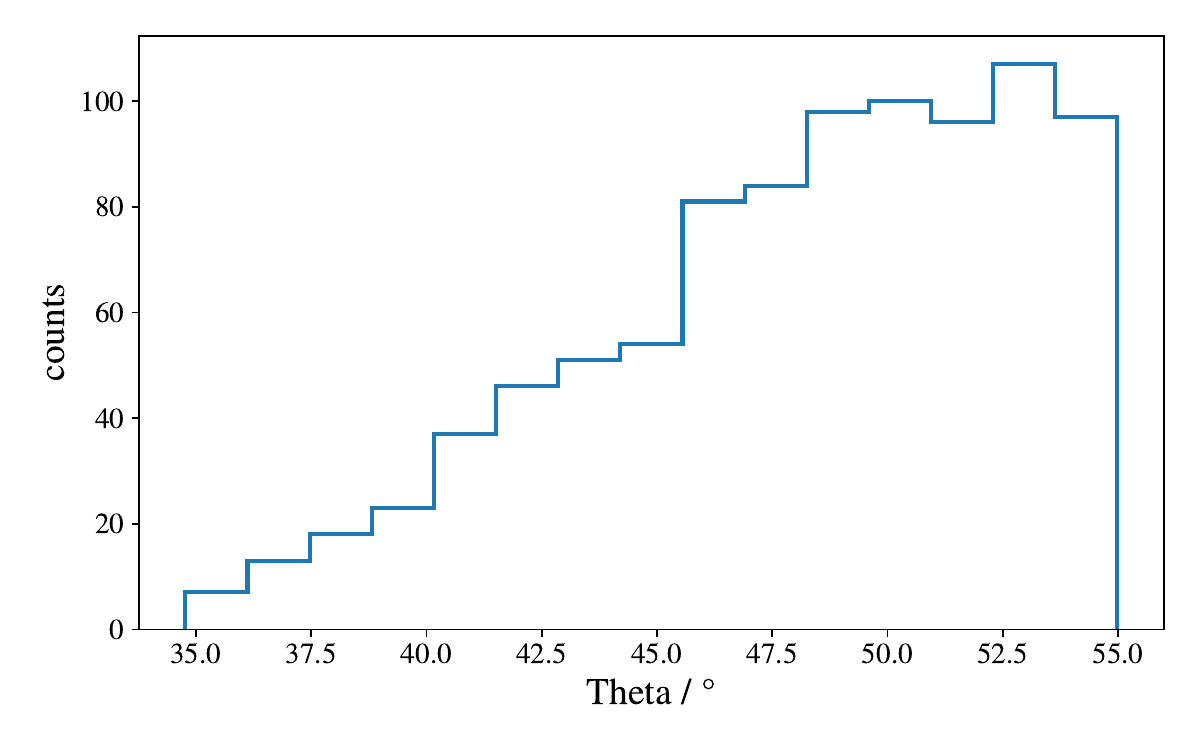}
    \subcaption{Zenith angles}
  \end{subfigure}\hfill
  \begin{subfigure}[c]{0.5\textwidth}
    \center
    \includegraphics[width=0.95\textwidth]{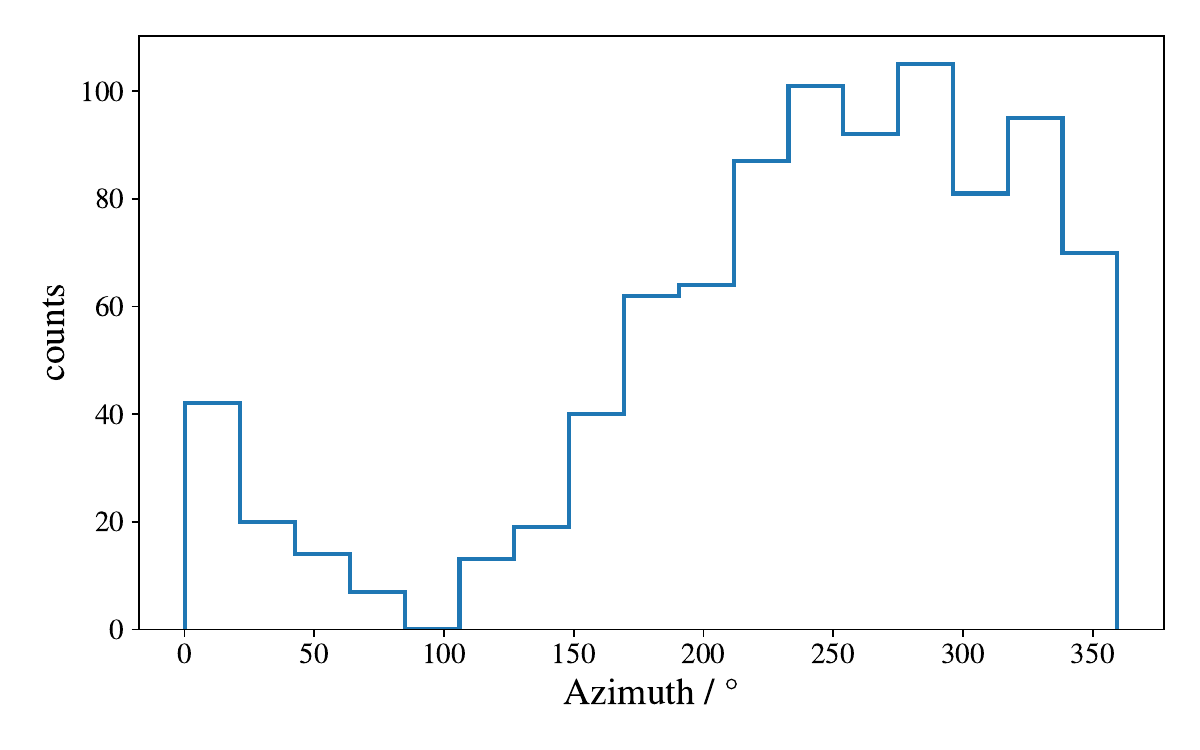}
    \subcaption{Azimuth angles}
  \end{subfigure}
  \caption{Distributions of the SD-reconstructed CR energies, shower cores, zenith angles and azimuth angles of the hybrid SD-AERA events.}
  \label{fig:dataset}
\end{figure}

The distribution of several variables of the dataset is shown in Fig.\ \ref{fig:dataset}. The distribution of energies reconstructed by the SD exhibit a steep decrease from the minimum energy towards higher energies, which is generally expected. The most energetic hybrid events lie around \SI{e19}{eV}. The reconstructed shower cores fall inside the part of the externally triggered AERA station array that has a spacing dense enough to meet the 5-signal-station requirement. (The more sparsely installed stations in the southern part of AERA are used to detect more inclined showers.) Fig.\ \ref{fig:dataset} b) shows an uneven distribution of the shower cores which is not unexpected due to the intrinsic inhomogeneity of AERA as an engineering array. The distributions of the zenith and azimuth angles of the reconstructed shower axes look as expected: most showers have zenith angles around \SI{50}{\degree} and the North-South asymmetry in the azimuth distribution, which arises from the orientation of the showers with respect to the Earth's magnetic field, is clearly recognizable.

\section{Discussion and conclusions}
\label{sec:7_Discussion}

The presented analysis method will allow us to make an independent cross-check of the CR energy scale of the Pierre Auger Observatory that is defined by the fluorescence detection method by probing its agreement with the radio energy scale. At the basis of the analysis, we use the radiation energy emitted by air showers which were detected in hybrid mode by both the SD and AERA. Concretely, we compare the reconstructions of measured events and of matched simulations. Several ingredients that were developed in the past years with considerable effort enable a high quality of this analysis:

\begin{itemize}
    \item \textbf{Monte-Carlo simulation of air-shower radio emission:} [Status: ready]
    We simulate the air showers using CORSIKA and CoREAS for the particle cascade and the radio emission, respectively. The simulation of air-shower radio signals has matured well and is in a good state. There is good agreement between the signals predicted by the two major codes CoREAS and ZHAireS at a level of about 3\% on the energy scale \cite{Gottowik_2018}. This validation is important as the simulation of the radio emission is at the foundations of the radio energy scale.

    \item \textbf{Event simulations:} [Status: ready]
    Each simulation is matched as realistically as possible to the measured event with the usage of GDAS atmospheres from the location and time of the event and with settings that provide accurate results. Measured noise taken from periodic triggers close to the time of the event is added to the simulated signals individually for each station to ensure realistic conditions in the reconstruction.

    \item \textbf{AERA detector calibration:} [Status: ready]
    In the event reconstruction, we use an accurate detector description of AERA, including lab-measured signal responses of the electronics in the readout chain and simulated antenna patterns, which in the case of the LPDA antenna was rescaled with data from a drone calibration campaign. The antenna patterns were cross-checked with the sidereal evolution of the Galactic background. An accurate absolute calibration is provided by a full-fledged Galactic calibration with a systematic uncertainty on the CR energy scale of ~6\%.

    \item \textbf{Radio signal reconstruction:} [Status: ready but with possible improvements]
    Our radio event reconstruction is composed of several ingredients that have been developed carefully over the past years with extensive validation. 
    In the near future, we plan to employ an improved station-signal estimation method that is derived from statistical considerations and is expected to perform better than our currently used method \cite{Martinelli_2024}.

    \item \textbf{Hybrid dataset:} [Status: almost ready]
    We rely on a dataset of CR air-shower events measured in hybrid mode by both the SD and AERA. With strict selection cuts applied on both detector arrays, we obtain a set of 912 events which is unprecedented statistics for the radio detection technique at such a high quality level.

\end{itemize}

The long-time experience of CR radio detection at the Pierre Auger Observatory with AERA and many preceding studies and efforts allow us to perform the presented analysis aiming to cross-check the established CR energy scale of the Pierre Auger Observatory with an accuracy that is competitive with that of the fluorescence technique. Looking ahead, we will be able to perform the analysis as well using the newly deployed AugerPrime RD, expanding it to the highest energies and largest inclinations.

{
\acknowledgments
This research was supported by the Bundesministerium für Bildung und Forschung (BMBF) under the contract 05A23VK4 and by the Hermann von Helmholtz-Gemeinschaft Deutscher Forschungszentren e.V. through the Helmholtz International Research School for Astroparticle Physics (HIRSAP), part of Initiative and Networking Fund (grant number HIRS-0009).
}

\setlength{\bibsep}{0pt plus 0.3ex}
\bibliographystyle{MyJHEP}
{\small
\bibliography{refsm}
}


\end{document}